\documentclass[prl,aps,showpacs,twocolumn,amsmath,amssymb]{revtex4}

\usepackage{graphicx}
\usepackage{dcolumn}
\usepackage{bm}
\usepackage{natbib}
\usepackage{color}

\begin{document}


\title{Laser-driven plasma waves in capillary tubes}

\author{}
\author{
F.~Wojda$^{1}$, K.~Cassou$^{1}$, G.~Genoud$^{2}$, M.~Burza$^{2}$,
Y.~Glinec$^{2}$, O.~Lundh$^{2}$, A.~Persson$^{2}$,
 G. Vieux$^3$, E. Brunetti$^3$, R.P. Shanks$^3$, D. Jaroszynski$^3$, N.~E.~Andreev$^{4}$,
C.-G.~Wahlstr\"om$^{2}$,  and~B.~Cros$^{1}$ }
 \email{brigitte.cros@u-psud.fr}

\address{%
\vskip 0.30cm \centerline{$^1$Laboratoire Physique Gaz et Plasmas
CNRS - Universit\'e Paris-Sud 11, F-91405 Orsay Cedex, France\\}
\centerline{$^2$Department of Physics, Lund University, P.O. Box
118, S-22100 Lund, Sweden\\}  \centerline{$^3$Department of Physics,
 University of Strathclyde,
Glasgow G4 0NG, UK \\
}\centerline{$^4$Joint Institute for High Temperatures, Russian
Academy of Sciences, Moscow 125412, Russia\\}
}%

\date{\today}

\begin{abstract}

The excitation of plasma waves over a length of up to $8$
centimeters is, for the first time, demonstrated using laser guiding
of intense laser pulses through hydrogen filled glass capillary
tubes. The plasma waves are diagnosed by spectral analysis of the
transmitted laser radiation. The dependence of the spectral
redshift, measured as a function of filling pressure, capillary tube
length and incident laser energy, is in excellent agreement with
simulation results. The longitudinal accelerating field  inferred
from the simulations is in the range $1-10\,$ GV/m.



\end{abstract}

\maketitle

Electrons accelerated in  laser wakefield accelerators
(LWFAs)\cite{tajima1979,Sprangle1988,Esarey1996} acquire momentum
from the electrostatic fields of a plasma density wave or wake,
excited by an intense laser pulse passing through  plasma. As they
can achieve ultra-relativistic energies over very short distances,
they have attracted great interest as novel routes to a new
generation of ultra-compact accelerators, which could be relevant to
a wide range of applications, including high-energy physics and
compact free-electron lasers. At high laser
intensities large amplitude plasma waves with longitudinal electric
fields, up to hundreds of GV/m, are excited, which are
several orders of magnitude higher than possible in conventional
accelerators. When electrons are injected into the accelerating
regions of such plasma waves, either from an external source \cite{Amiranoff1998} or by
self-trapping of plasma electrons by non-linear effects, they have
been observed to reach energies up to hundreds of MeV over acceleration
distances of only a few mm
\cite{Mangles2004a,Geddes2004a,Faure2004,Leemans2006}.
However, even with fields as high as 100 GV/m, the acceleration
distance must be extended to tens of cm in order to reach electron
energies of, say, tens of GeV. This is a range not yet realized
experimentally.

There are two main requirements that must be fulfilled before a
viable multi-GeV LWFA can be realized. First, the intense laser
pulse driving the plasma wave must be guided over the full length of
the accelerator. Second, the dephasing length, which is the distance
over which the accelerating electrons outrun the accelerating region
of the plasma wave, must be at least as long as the accelerating
medium. To date, most experiments have been performed with a
few mm long gas jets, where self-guiding due to relativistic
self-focusing provides a simple guiding solution. However, the
threshold power for self-focusing is, for a given laser wavelength,
inversely proportional to the plasma density, thus requiring a
minimum density for a given laser. As an example, for a laser with
10 TW peak power and 800 nm wavelength, this density is $n_e\approx
3 \times 10^{18}\,$cm$^{-3}$. An alternative approach  is to guide
the laser in  a preformed plasma channel produced by an electric
discharge in a gas filled capillary tube. In this way guiding up to
several centimeter long plasma waveguide has been demonstrated
\cite{Spence2000}. However, this approach is limited to plasma
densities in excess of $n_e \sim10^{18}\,$cm$^{-3}$.
 Long dephasing
lengths require very low plasma densities \cite{Esarey1996}, which
is clearly in conflict with the density requirements for
relativistic self-guiding and guiding by discharge-formed plasma
channels.

A method of guiding the intense laser pulse over long distance,
while allowing very low plasma density, is thus
required. This can be realized using capillary tubes as waveguides
\cite{Dorchies1999} where guiding is achieved by reflection from the
walls. Laser guiding through gas-filled capillary tubes  allows for
smooth transverse profile in monomode propagation \cite{Cros2002}
and minimum attenuation of the laser pulse through refraction
losses. As a first step in developing the waveguide as a medium for
a wakefield accelerator, we have characterized the properties of the
plasma wave in the moderately non-linear regime over several
centimeters. Working in the linear or moderately non-linear regime
allows control of the amplitude of the plasma waves while providing
a focal spot of sufficient dimensions to effectively create quasi
$1$-dimensional, longitudinal oscillations as shown by theoretical
studies \cite{Andreev2002a, Andreev2002b}. This regime enables controlled and
reproducible injection of an external electron bunch into the
accelerating field, which provides a means of attaining high
stability and reproducibility for future staging of several LWFA
stages.

In this letter, we report on the first observation of plasma waves
excited by a guided high-intensity laser pulse inside hydrogen
filled capillary tubes with lengths up to $8\,$cm.  The plasma wave
excitation is diagnosed using a method based on spectral
modifications of the laser pulse, due to local spatio-temporal
variations of the density of the plasma \cite{Andreev2005}.

\begin{figure}[h]
\includegraphics[width=7.8cm]{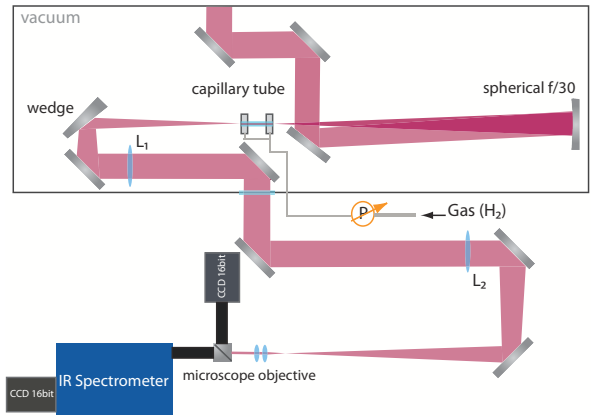}
\caption{(color online) Schematic view of the experiment setup. Elements
within the gray line box are under vacuum.} \label{fig:fig1}
\end{figure}

An experiment was performed using the high intensity Ti:Sapphire
laser system at the Lund Laser Centre which  delivers up to 40 TW
onto the target, with a full width
at half maximum (FWHM) pulse duration down to $\tau_L = 35\,$fs.
The set-up is shown schematically in Fig. \ref{fig:fig1}. The
$30\,$mm diameter beam was focused with an $f=1.5\,$m spherical
mirror at the entrance of a capillary tube. A deformable mirror was
placed after the compressor (not shown) to correct the main
aberrations of the phase front in the focal plane.

The laser beam transmitted through the capillary tube was attenuated
by reflecting it on an optically flat glass wedge and then
collimated by an achromatic $f_1=1\,$m lens ($L_1$), which could be
translated in  vacuum along the beam axis to collect light either
from the focal plane (i.e. with the capillary tube removed) or from
the exit plane of the tube. The beam was then focused by a
$f_2=1\,$m achromatic lens ($L_2$) and magnified 10 $\times$ by a
microscope objective. The beam was split in two parts. Focal spot or
capillary tube output images were recorded by a 16-bit
charged-couple-device (CCD). The transmitted part was sent to a
visible to near infrared imaging spectrometer equipped with a 16-bit
CCD camera. The spectral resolution was $0.1\,$nm.

Glass capillary tubes with inner radius $r_c=50\,\mu$m and length
varying between $1.2\,$ and $8.1\,$cm were used. Hydrogen gas flowed
into the tubes through two thin ($\sim100\,\mu$m) slits located
between $2.5\,$mm and $5\,$mm from each end of the tube. The filling
pressure was varied between $0\,$ and $70\,$mbar. Each capillary
tube could be used for at least one hundred laser shots, when the laser
beam remained well centered at the capillary entrance. Pointing
variations due to thermal drifts and mechanical vibrations were
therefore minimized or compensated for. Laser guiding at input
intensities up to $10^{18}\, $W/cm$^{2}$ was achieved with more
than $90 \%$ energy transmission in evacuated or hydrogen filled gas
tubes up to 8~cm long.

For the data presented here, in order to investigate the moderately
non linear regime, the input intensity was kept lower than $3 \times
10^{17}\,$W/cm$^{2}$. The laser pulse duration was $\tau_L=45 \pm5
\,$fs and the associated bandwidth approximately $25\,$nm (FWHM);
each pulse had a small negative linear chirp ($-550\,$fs$^2$), i.e. short wavelengths
preceded longer wavelengths, and the center wavelength was 786~nm.
The energy distribution in the focal plane exhibited an Airy-like
pattern with a radius at first minimum of $r_0= 40 \pm 5\, \mu$m.


Spectra of the laser light transmitted through gas filled capillary tubes
 exhibit blue and red broadening. In the range of parameters relevant
to this experiment, spectral modifications of the wake-driving laser
pulse, after propagating in the plasma over a large distance, are
mainly related to changes in the index of refraction of the plasma
during the creation of the plasma wave. The front of the laser pulse
creates an increase in electron density, leading to a blue-shift at
the front of the pulse, while the rear of the pulse creates a
decrease of electron density with larger amplitude, and thus a
red-shift of the spectrum. This effect has been proposed \cite{Andreev2005} to
determine the amplitude of the electron plasma density perturbation
in the linear or moderately nonlinear regime.

An averaged wavelength shift, $\Delta \lambda (l)$, is calculated
from the experimentally measured spectra, $S(\lambda, l)$, at the
exit of a capillary of length $l$, and is defined as
\begin{equation}
\Delta \lambda (l) = \frac{\int_0 ^{\infty} \lambda S(\lambda, l)
d\lambda}{ \int_0 ^{\infty} S (\lambda, l) d\lambda } - \lambda _L
\end{equation}
where  $ \lambda _L= \left[ \int_0 ^{\infty}  S(\lambda,l=0)
d\lambda\right]^{-1} \int_0 ^{\infty} \lambda S(\lambda,l=0) \simeq 2
\pi c / \omega _L$  is the center wavelength of the incident laser
pulse in vacuum, and $\omega_L$ is the laser
frequency. For an underdense plasma and laser intensity well above
the ionization threshold, when the blue-shift due to gas ionization
inside the interaction volume $V$ can be neglected and the
wavelength shift given by Eq.~(1) remains small compared to
$\lambda_L$, this shift is directly related  to the energy of the
plasma wave electric field, $E_p$, excited in the
plasma~\cite{Andreev2005}:
\begin{equation}
\frac{\Delta \lambda (l)}{\lambda _L} \simeq \frac{1}{16 \pi
\mathcal{E}_{out}} \int_V E_p ^2 \,dV \, ,
\end{equation}
where $\mathcal{E}_{out}$ is the total energy of the transmitted pulse. For
monomode propagation of a laser pulse with Gaussian time envelope,
generating a wakefield in the weakly non linear regime, the
wavelength shift can be expressed analytically. For small energy
losses, it is proportional to the peak laser intensity on the
capillary axis, to the length of the capillary and exhibits a
resonant-like dependence on gas pressure described by the function
$D(\Omega)$, as:
\begin{equation}
\Delta \lambda (l) \simeq
 \left[0.178 + 1.378 \frac{c^2}{(\omega_p
r_c)^2}\right] \left( \frac{\omega_p}{\omega_L}\right)^3 a_L^2
D(\Omega) l \, ,
\end{equation}
where $a_L=e E_L/m_e\omega_L c$ is the normalized amplitude of  laser electric field $E_L$, $D(\Omega)
= \Omega \exp(-\Omega^2/4)$ with $\Omega = \omega_p \tau_L/\sqrt{2
\ln 2}$,  and $\omega_p=\sqrt{n_e e^2/\epsilon_0 m_e}$ is the electron plasma frequency .

\begin{figure}[h]
\includegraphics[width=7.8cm]{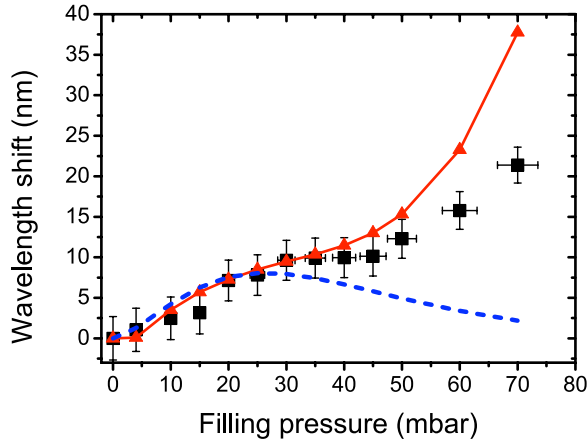}
\caption{(color online). Wavelength shift as a function of hydrogen
H$_2$ filling pressure at the exit of a $7.1\,$cm -long capillary
tube obtained from experimental data (black squares); simulation
results (red triangle curve) with $(\tau_L, \mathcal{E}_L)= (51\, \rm{fs},
0.12\,\rm{J})$ and a negative chirp; analytical result from eq.~(3)
for the same parameters (dashed blue line).} \label{fig:fig2}
\end{figure}

 The value of the wavelength shift $ \Delta \lambda $ obtained from
measured spectra is plotted as a series of black squares in
Fig.~\ref{fig:fig2} for a tube of length
 $7.1\,$cm  as a function of filling pressure.

Nonlinear laser pulse propagation in the gas filled capillary tube,
including optical field ionization of gas, wakefield generation and
the self-consistent laser pulse spectrum modification were simulated
numerically using the code described in \cite{Andreev2002a}. The
parameters used at the capillary entrance are $\tau_L=51\,$fs with a
negative chirp, and an incident laser energy  $\mathcal{E}_L=0.12\,$J, with a
radial profile corresponding to the one measured in the focal plane
in vacuum, averaged over the angle. Simulation results are plotted
as a red triangles curve in Fig.~2. For comparison, the analytical
behavior given by Eq.~(3) is plotted as a blue dashed curve, for the
same ($\tau_L , \mathcal{E}_L$). The simulated wavelength shifts fit closely
the experimentally measured ones  up to a filling pressure of 50
mbar. Analytical curve, measured and modeled wavelength shifts have
the same behavior up to the linear resonant pressure, 25~mbar ($n_e
\simeq 1.2 \times 10^{18}\,$cm$^{-3}$). For pressures higher than
25~mbar, the experimental data and simulation exhibit a larger red
shift than the analytical prediction. The analysis of the pulse
evolution in the simulation shows that a steepening of the laser
pulse front edge, due to propagation in the ionizing
gas~\cite{AndreevJETP03} occurs, followed in time by pulse
self-modulation~\cite{Andreev2002a}. A steepened pulse front is more
efficient than a Gaussian time envelope to generate the wakefield at
pressures larger than the resonant one. The values of the wavelength
shift in the simulation are larger than the measured ones for
pressures above 50~mbar. This behavior is related to the assumption
of cylindrical symmetry used in the simulation, which leads to more
pronounced non-linear effects including laser pulse shortening 
~\cite{Andreev2002a, SkobelevJETPlett09}. 

Figure 3 shows  examples of the spectra of the laser pulse measured
(black lines) a) in the focal plane in vacuum, and at the output of
the 7.1 cm capillary tube and the corresponding simulated spectra
(dashed red lines) for filling pressures of b) 30~mbar and c)
40~mbar; two different shots are shown for each pressure.
\begin{figure}[h]
\includegraphics[width=7.8cm]{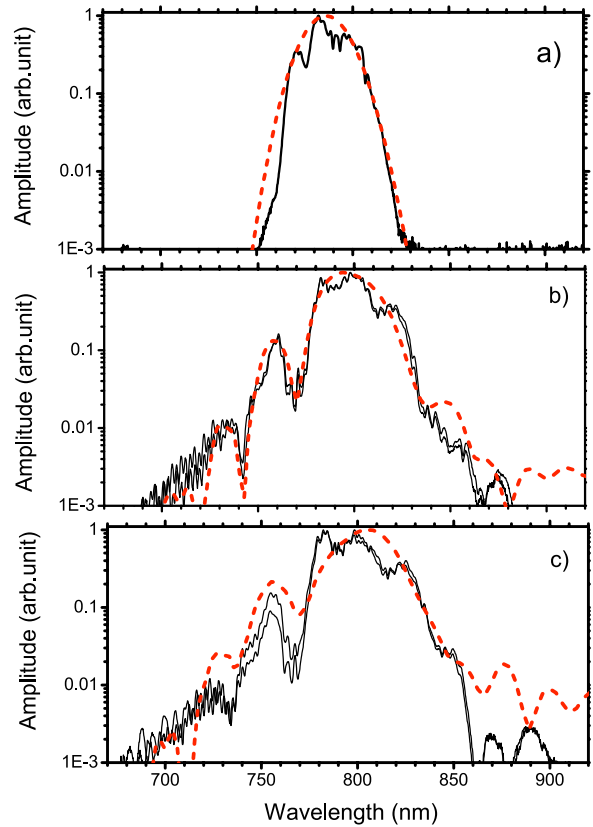}
\caption{(color online). Spectra measured (black lines) a) in the
focal plane in vacuum and at the output of the 7.1 cm capillary tube
for filling pressures of b) 30 mbar and c) 40mbar; two shots are
shown for each pressure; corresponding simulated spectra (red dashed
line). } \label{fig:fig3}
\end{figure}
All spectra are normalized to their maximum amplitude and integrated
over the radial coordinate. The spectral modifications in the
experiment and the simulations are in excellent agreement over a
large range of amplitude. This agreement on the detailed structure
of the spectra confirms
 the one, shown in Fig.~2, achieved on the wavelength shift which is an integrated and averaged
 quantity.


The wavelength shift measured at the output of the capillary tubes
as a function of the tube length  is plotted in Fig.~4
\begin{figure}[h]
\includegraphics[width=7.8cm]{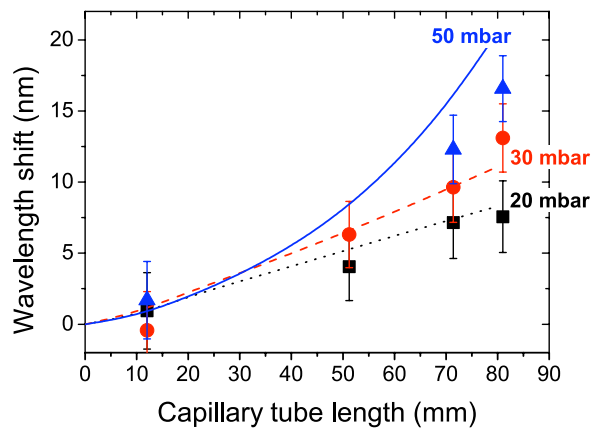}
\caption{(color online). Wavelength shift measured at the output of
the capillary tubes as a function of the tube length  for filling
pressures $20\,$mbar (black square), $30\,$mbar (red dots) and
$50\,$mbar (blue triangles) and the corresponding simulated shifts
(dotted, dashed and solid lines).} \label{fig:fig4}
\end{figure}
for different filling pressures  as well as the corresponding
simulated results. A linear behavior of the wavelength shift as a
function of length is observed at 20~mbar. The fit of  experimental
data by simulation results demonstrates that the plasma wave is
excited over a length as long as $8\,$cm. As the pressure is
increased, the non linear laser pulse evolution is amplified with
the propagation length leading to a larger plasma wave amplitude.

The wavelength shift as a function of input laser energy is shown in
Fig.~5,
\begin{figure}[h]
\includegraphics[width=7.8cm]{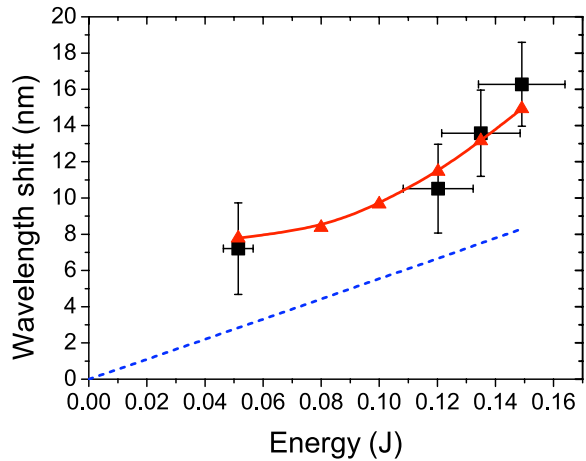}
\caption{(color online). Wavelength shift as a function of input
energy, at the exit of a $7.1\,$cm long capillary tube for a filling
pressure of $40\,$mbar obtained from experimental data (black
squares), simulation results (red triangle curve), and analytical
prediction (blue dashed curve).} \label{fig:fig5}
\end{figure}
at the exit of a $7.1\,$cm long capillary tube for a filling
pressure of $40\,$mbar,  obtained from experimental data, simulation
results and analytical prediction. For $\mathcal{E}_L>0.1\,$J, the growth of
the wavelength shift is faster than the linear one predicted
analytically and the experimental and numerical results are in
excellent agreement. For lower energy ($\simeq 0.05\, $J), gas
ionization occurs closer to the maximum of the pulse and, combined
with  radial structure, leads to a more pronounced steepening of the
front edge of the pulse and increased wakefield amplitude and
wavelength shift compared to the analytical prediction, eq. (3).

 In conclusion, we have generated and characterized, for
the first time, a laser-driven plasma wave in the moderately non
linear regime over a distance as long as $8\,$cm inside dielectric
capillary tubes.  An excellent agreement is found between the
measured wavelength shift and the results from simulations as
concerns pressure, length and energy dependence. In the linear and
weakly non linear regimes, this diagnostic provides a robust and
 reproducible measurement of the plasma wave amplitude over a long distance.
The value of the longitudinal accelerating field in the plasma
obtained from the simulation is in the range 1-10 GV/m. The average
product of gradient and length achieved in this experiment is of the
order of 0.4~GV at a pressure of 50~mbar; it could be increased to
several~GV by extending the length and diameter of the capillary
tube with higher laser energy.

\begin{acknowledgments}
We acknowledge the support of the European Community -New and
Emerging Science and Technology Activity under the FP6 "Structuring
the European Research Area" programme (project EuroLEAP, contract
028514), -Marie Curie Early Stage Training Site MAXLAS
(MEST-CT-2005-020356). This work was also supported by the Swedish
Research Council, the Knut and Alice Wallenberg Foundation, the EU Access to Research Infrastructures activity (contract RII3-CT-2003-506350, Laserlab Europe), the Russian
Foundation for Basic Research (project 07-02-92160), and the EPSRC UK.
\end{acknowledgments}

\bibliography{apssamb-1}

\end{document}